\documentclass{article}
\title{\bf Time-dependent spherically-symmetric\protect\\ 5-D vacuum solutions.}
\author{ Sergey S. Kokarev\thanks{e-mail:
logos-center@mail.ru}\ and Vladimir G. Krechet
}
\date{Regional Scientific Educational Center "Logos"}

\begin{document}
\maketitle












\begin{abstract}
Vacuum 5-D Einstein equations with spherical symmetry and t-dependence
are considered. For the case of separating variables several classes of
exact solutions are obtained. Effective matter, induced by geometrical
scalar field $\varphi=\sqrt{-G_{55}}$ is analyzed.
\end{abstract}




%

\section{Introduction}

Introducing of extradimension into GR theory endows solutions
to multidimensional Einstein equations   qualitatively  new properties
in comparison with their 4-D analogous.
So, for example, in 4-D GR the following result is well known ({\it Birkhoff
theorem}): under definite conditions Shwarzschild solution is unique
spherically-symmetric solution of vacuum 4-D Einstein equation \cite{Shm}.
Consequence of this theorem is absence of spherically-symmetric
nonstationary vacuum solutions.

In 5-D GR probably alleviate variant of {\it Birkhoff
theorem} take place (see also \cite{bron}):
static spherically-symmetric solution to  5-D vacuum Einstein equation
is unique (Kramer's solution), but there is variety of nonstationary
spherically-symmetric vacuum solutions.

The aim of present paper is obtaining and analysis of such time-dependent
Kaluza-Klein solitons (the term of Wesson \cite{wes1,wes2}), for cases of
separating
variables. This problem have been stated in \cite{wes2}, and have been
solved  for time-part equations.

First half  of present  article gives  mathematical part of our
investigation:
it contains basic equations and it's solutions (if they have been found) and
second half devoted to physical analysis: what effective matter obtained
solutions induce?\cite{wes3,kok1}

\section{Exact solutions of vacuum
5-D Einstein equations}\label{sol}

The starting metric without loss of generality can be written in the
following form:
\begin{equation}\label{sphmetr}
ds^{2}=e^{\nu(r,t)}dt^{2}-e^{\lambda(r,t)}dr^{2}-e^{\mu(r,t)}d\Omega^{2}
-e^{2\phi(r,t)}(dx^{5})^{2}
\end{equation}
Nonzero Cristoffel's symbols are:
\[
\begin{array}{ll}
\Gamma^{0}_{00}={\displaystyle\frac{\dot{\nu}}{2}};&\
\Gamma^{1}_{11}={\displaystyle\frac{\lambda^{\prime}}{2}};\\
\\
\Gamma^{0}_{11}={\displaystyle\frac{\dot{\lambda}}{2}}e^{\lambda-\nu};&\
\Gamma^{0}_{22}={\displaystyle\frac{\dot{\mu}}{2}}e^{\mu-\nu};\\
\\
\Gamma^{0}_{33}={\displaystyle\frac{\dot{\mu}}{2}}e^{\mu-\nu}\sin^{2}\theta;&\
\Gamma^{1}_{00}={\displaystyle\frac{\nu^{\prime}}{2}}e^{\nu-\lambda};\\
\\
\Gamma^{1}_{22}=-{\displaystyle\frac{1}{2}}\mu^{\prime} e^{\mu-\lambda};&\
\Gamma^{1}_{33}=-{\displaystyle\frac{1}{2}}\mu^{\prime} e^{\mu-\lambda}\sin^{2}\theta;\\
\\
\Gamma^{2}_{33}=-\sin\theta\cos\theta;&\
\Gamma^{0}_{01}={\displaystyle\frac{\nu^{\prime}}{2}};\\
\\
\Gamma^{1}_{10}={\displaystyle\frac{\dot{\lambda}}{2}};&\
\Gamma^{2}_{20}={\displaystyle\frac{\dot{\mu}}{2}};\\
\\
\Gamma^{2}_{21}={\displaystyle\frac{\mu^{\prime}}{2}};&\
\Gamma^{3}_{30}={\displaystyle\frac{\dot{\mu}}{2}};\\
\\
\Gamma^{3}_{31}={\displaystyle\frac{\mu^{\prime}}{2}};&\
\Gamma^{3}_{32}=\cot\theta;\\
\\
\Gamma^{0}_{55}=\dot{\phi}e^{2\phi-\nu};&\
\Gamma^{1}_{55}=-\phi^{\prime} e^{2\phi-\lambda};\\
\\
\Gamma^{5}_{50}=\dot{\phi};&\
\Gamma^{5}_{51}=\phi^{\prime}
\end{array}
\]
5-D Einstein equations $R_{AB}=0$ for the metric (\ref{sphmetr}) have the
following kind:
\begin{equation}\label{00}
R_{00}=\frac{e^{\nu-\lambda}}{2}(\nu''+\frac{\nu'^{2}}{2}-
\frac{\nu'\lambda'}{2}+\nu'\mu'+\phi'\nu')
\end{equation}
\[
-\ddot{\mu}-\ddot{\phi}-\frac{\ddot{\lambda}}{2}+
\frac{\dot{\nu}}{2}(\frac{\dot{\lambda}}{2}+\dot{\mu}+\dot{\phi})-
\frac{{\dot{\lambda}}^{2}}{4}-\frac{\dot{\mu}^{2}}{2}-\dot{\phi}^{2}=0;
\]

\begin{equation}\label{11}
R_{11}=\frac{e^{\lambda-\nu}}{2}(\ddot{\lambda}+\frac{\dot{\lambda}^{2}}{2}-
\frac{\dot{\lambda}\dot{\nu}}{2}+\dot{\lambda}\dot{\mu}+\dot{\phi}\dot{\lambda})
-\mu''-\phi''
\end{equation}
\[
-\frac{\nu''}{2}+\frac{\lambda'}{2}
(\frac{\nu'}{2}+\mu'+\phi')-\frac{\nu'^{2}}{4}-
\frac{\mu'^{2}}{2}-\phi'^{2}=0;
\]

\begin{equation}\label{22}
R_{22}=\frac{e^{\mu-\nu}}{2}(\ddot{\mu}+\dot{\mu}^{2}-\frac{\dot{\mu}\dot{\nu}}{2}+
\frac{\dot{\mu}\dot{\lambda}}{2}+\dot{\phi}\dot{\mu})
-\frac{e^{\mu-\lambda}}{2}
(\mu''+\mu'^{2}-\frac{\mu'\lambda'}{2}+
\frac{\mu'\nu'}{2}+\mu'\phi')+1=0;
\end{equation}

\begin{equation}\label{33}
R_{33}=R_{22}\sin^{2}\theta;
\end{equation}

\begin{equation}\label{55}
R_{55}=e^{2\phi-\nu}(\ddot{\phi}+\dot{\phi}^{2}-\frac{\dot{\nu}\dot{\phi}}{2}+
\frac{\dot{\lambda}\dot{\phi}}{2}+\dot{\phi}\dot{\mu})
-e^{2\phi-\lambda}
(\phi''+{\phi'}^{2}-\frac{\lambda'\phi'}{2}+
\frac{\nu'\phi'}{2}+\mu'\phi')=0;
\end{equation}

\begin{equation}\label{01}
R_{01}=-\dot{\mu}'-\dot{\phi}'+\frac{\nu'}{2}(\dot{\mu}+
\dot{\phi})+\frac{\dot{\lambda}}{2}(\mu'+\phi')-
\frac{\mu'\dot{\mu}}{2}
-\phi'\dot{\phi}=0.
\end{equation}

Case of separating variables, which is within of our interest,
is characterized by the special kind of metric functions:
\[
\nu(r,t)=\nu_{1}(t)+\nu_{2}(r);\ \ \lambda(r,t)=\lambda_{1}(t)+
\lambda_{2}(r);
\]
\[
\mu(r,t)=\mu_{1}(t)+\mu_{2}(r);\ \ \phi(r,t)=\phi_{1}(t)+\phi_{2}(r),
\]
In this case system (\ref{00})--(\ref{01}) takes the following form:
\begin{equation}\label{dt00}
e^{\lambda_{1}-\nu_{1}}(\ddot\mu_{1}+\frac{\ddot\lambda_{1}}{2}-
\frac{\dot\nu_{1}}{2}\left(\frac{\dot\lambda_{1}}{2}+\dot\mu_{1}+
\dot\phi_{1}\right)
+\frac{\dot\lambda_{1}^{2}}{4}+\frac{\dot\mu_{1}^{2}}{2}+
\ddot\phi_{1}+\dot\phi_{1}^{2})=\alpha;
\end{equation}

\begin{equation}\label{dr00}
\frac{e^{\nu_{2}-\lambda_{2}}}{2}(\nu_{2}''+\frac{\nu_{2}'^{2}}{2}-
\frac{\nu_{2}'\lambda_{2}'}{2}+\nu_{2}'\mu_{2}'+\nu_{2}'\phi_{2}')=\alpha;
\end{equation}

\begin{equation}\label{dt11}
\frac{e^{\lambda_{1}-\nu_{1}}}{2}(\ddot\lambda_{1}+\frac{\dot\lambda_{1}^{2}}{2}-
\frac{\dot\lambda_{1}\dot\nu_{1}}{2}+\dot\lambda_{1}\dot\mu_{1}+
\dot\lambda_{1}\dot\phi_{1})=\beta;
\end{equation}
\begin{equation}\label{dr11}
e^{\nu_{2}-\lambda_{2}}(\mu_{2}''+\frac{\mu_{2}'^{2}}{2}+\frac{\nu_{2}''}{2}+
\frac{\nu_{2}'^{2}}{4}
-\frac{\lambda_{2}'}{2}(\frac{\nu_{2}'}{2}+\mu_{2}'+
\phi_{2}')
+\phi_{2}''+\phi_{2}'^{2})=\beta;
\end{equation}

\begin{equation}\label{dtr22}
\frac{e^{\lambda_{1}-\nu_{1}}}{2}(\ddot\mu_{1}+\dot\mu_{1}^{2}-
\frac{\dot\mu_{1}\dot\nu_{1}}{2}+\frac{\dot\mu_{1}\dot\lambda_{1}}{2}+
\dot\mu_{1}\dot\phi_{1})
-\frac{e^{\nu_{2}-\lambda_{2}}}{2}(\mu_{2}''+\mu_{2}'^{2}-
\frac{\mu_{2}'\lambda_{2}'}{2}+\frac{\mu_{2}'\nu_{2}'}{2}+\mu_{2}'\phi_{2}')
\end{equation}
\[
+e^{\nu_{2}-\mu_{2}+\lambda_{1}-\mu_{1}}=0;
\]

\begin{equation}\label{dt55}
e^{\lambda_{1}-\nu_{1}}(\ddot\phi_{1}+\dot\phi_{1}^{2}-
\frac{\dot\nu_{1}\dot\phi_{1}}{2}+\frac{\dot\lambda_{1}\dot\phi_{1}}{2}+
\dot\mu_{1}\dot\phi_{1})=\gamma;
\end{equation}
\begin{equation}\label{dr55}
e^{\nu_{2}-\lambda_{2}}(\phi_{2}''+\phi_{2}'^{2}-
\frac{\lambda_{2}'\phi_{2}'}{2}+\frac{\nu_{2}'\phi_{2}'}{2}+
\mu_{2}'\phi_{2}')=\gamma;
\end{equation}

\begin{equation}\label{dtr01}
\frac{\nu_{2}'}{2}(\dot\mu_{1}+\dot\phi_{1})+\frac{\dot\lambda_{1}}{2}
(\mu_{2}'+\phi_{2}')-\frac{\mu_{2}'\dot\mu_{1}}{2}-\phi_{2}'\dot\phi_{1}=0.
\end{equation}

Taking derivative of eq. (\ref{dtr22}) with respect to $r$ and then with respect to  $t$ we get:
\begin{equation}\label{cond1}
\nu_{2}=\mu_{2}+const\ \ \mbox{\rm or}\ \ \lambda_{1}=\mu_{1}+const.
\end{equation}

Let us at first consider the case, when both conditions (\ref{cond1})
are satisfied simultaneously.
Base system of equation (\ref{dr00})---(\ref{dtr01}) in this case takes the form:
\[
\frac{e^{\nu_{2}-\lambda_{2}}}{2}(\nu_{2}''+\frac{3}{2}\nu_{2}'^{2}-
\frac{\nu_{2}'\lambda_{2}'}{2}+\nu_{2}'\phi_{2}')=\alpha;
\]
\[
e^{\lambda_{1}-\nu_{1}}(\frac{3}{2}\ddot\lambda_{1}+
\frac{3}{4}\dot\lambda_{1}^{2}-
\frac{\dot\nu_{1}}{2}(\frac{3}{2}\dot\lambda_{1}+\dot\phi_{1})
+\ddot\phi_{1}+
\dot\phi_{1}^{2})=\alpha;
\]
\[
\frac{e^{\lambda_{1}-\nu_{1}}}{2}(\ddot\lambda_{1}+\frac{3}{2}\dot\lambda_{1}^{2}-
\frac{\dot\lambda_{1}\dot\nu_{1}}{2}+\dot\lambda_{1}\dot\phi_{1})=\beta;
\]
\[
e^{\nu_{2}-\lambda_{2}}(\frac{3}{2}\nu_{2}''+\frac{3}{4}\nu_{2}'^{2}-
\frac{\lambda_{2}'}{2}(\frac{3}{2}\nu_{2}'+\phi_{2}')
+\phi_{2}''+\phi_{2}'^{2})=\beta;
\]
\[
\frac{e^{\lambda_{1}-\nu_{1}}}{2}(\ddot\lambda_{1}+\frac{3}{2}\dot\lambda_{1}^{2}
-\frac{\dot\lambda_{1}\dot\nu_{1}}{2}+\dot\lambda_{1}\dot\phi_{1})
-\frac{e^{\nu_{2}-\lambda_{2}}}{2}(\nu_{2}''+\frac{3}{2}\nu_{2}'^{2}-
\frac{\nu_{2}'\lambda_{2}'}{2}+\nu_{2}'\phi_{2}')+AB=0;\longrightarrow
\]
\[
\beta-\alpha+AB=0;
\]
\[
e^{\lambda_{1}-\nu_{1}}(\ddot\phi_{1}+\dot\phi_{1}^{2}-
\frac{\dot\nu_{1}\dot\phi_{1}}{2}+\frac{3}{2}\dot\lambda_{1}\dot\phi_{1})=\gamma;
\]
\[
e^{\nu_{2}-\lambda_{2}}(\phi_{2}''+\phi_{2}'^{2}-\frac{\lambda_{2}'\phi_{2}'}{2}+
\frac{3}{2}\nu_{2}'\phi_{2}')=\gamma;
\]
\[
\frac{\nu_{2}'}{2}(\dot\phi_{1}+\dot\lambda_{1})+\phi_{2}'(\frac{\dot\lambda_{1}}{2}-\dot\phi_{1})=0;
\]
Here have been introduced following designations:
\[\mu_{1}=\lambda_{1}+\overline{\mu}_{1}\
(\overline{\mu}_{1}={const});\quad
\mu_{2}=\nu_{2}+\overline{\mu}_{2}\ (\overline{\mu}_{2}={const});
\]
\[
e^{-\overline{\mu}_{1}}=A;\ \
e^{-\overline{\mu}_{2}}=B.
\]

We shall not consider process of solution this system in details,
because of  its simplicity.  All solutions are exhausted
by the following ones:

general 1-parametric solution (A-solution)
\begin{equation}\label{gen1}
ds^{2}=r^{2}dt^{2}-\frac{u^{2}}{3}t^{2}dr^{2}-\frac{u^{2}}{3}\frac{r^{2}t^{2}}
{(3-u^{2})}d\Omega^{2}
\pm r^{{\scriptstyle 2(1+u)}}t^{\frac{{\scriptstyle 2(3+u)}}{{\scriptstyle u}}}(dx^{5})^{2},
\end{equation}
and particular solutions under $u\rightarrow0$:
\begin{equation}\label{u0}
ds^{2}=r^{2}dt^{2}-dr^{2}-\frac{1}{3}r^{2}d\Omega^{2}\pm r^{2}\sinh^{2}\sqrt{3}t
(dx^{5})^{2},
\end{equation}
and under $u\rightarrow\infty$:
\begin{equation}\label{uinf}
ds^{2}=dt^{2}-t^{2}dr^{2}+\frac{1}{3}t^{2}d\Omega^{2}\pm t^{2}\sinh^{2}
\sqrt{3}r(dx^{5})^{2}.
\end{equation}
Values of parameter $u=\pm\sqrt{3}$ --- singular, under which there is no
physical solutions. We note here, that solutions (\ref{u0}) и (\ref{uinf})
can be obtained from each other by interchanging $r\longleftrightarrow t$
and by inversing of signature of 4-D part of interval.
This is general property for considered symmetry: if we take
some solutions of the system (\ref{00})--(\ref{01}) and make in it
formal redesignations $r\longleftrightarrow t;\ \
\ \ \nu\longrightarrow\lambda+i\pi;\ \
\lambda\longrightarrow\nu+i\pi;\ \ \mu\longrightarrow\mu+i\pi;$
then in result we obtain new solutions of the original system.
This fact is a sequence of the symmetry of coordinates  $r$ and $t$ in
starting metric (\ref{00})--(\ref{01}) and in starting system of equations.

For the case $\mu_{1}=\lambda_{1}+\overline{\mu}_{1};\ \
e^{-\overline{\mu}_{1}}=A=const$
system (\ref{00})--(\ref{01}) take the following form:
\[
\frac{e^{\nu_{2}-\lambda_{2}}}{2}(\nu_{2}''+\frac{\nu_{2}'^{2}}{2}-
\frac{\nu_{2}'\lambda_{2}'}{2}+\nu_{2}'\mu_{2}'+\nu_{2}'\phi_{2}')=\alpha;
\]
\[
e^{\lambda_{1}-\nu_{1}}(\frac{3}{2}{\ddot\lambda}_{1}-\frac{\dot\nu_{1}}{2}
(\frac{3}{2}\dot{\lambda}_{1}+\dot\phi_{1})+\frac{3}{4}\dot\lambda_{1}^{2}
+\ddot{\phi}_{1}+\dot\phi_{1}^{2})=\alpha;
\]
\[
\frac{e^{\lambda_{1}-\nu_{1}}}{2}(\ddot\lambda_{1}+\frac{3}{2}\dot\lambda_{1}^{2}
-\frac{\dot\lambda_{1}\dot\nu_{1}}{2}+\dot\lambda_{1}\dot\phi_{1})=\beta;
\]
\[
e^{\nu_{2}-\lambda_{2}}(\mu_{2}''+\frac{\mu'^{2}}{2}+\frac{\nu''_{2}}{2}+
\frac{\nu_{2}'^{2}}{4}
-\frac{\lambda_{2}'}{2}(\frac{\nu_{2}'}{2}+\mu_{2}'+
\phi_{2}')+\phi_{2}''+\phi_{2}'^{2})=\beta;
\]
\[
\beta-\frac{e^{\nu_{2}-\lambda_{2}}}{2}(\mu_{2}''+\mu_{2}'^{2}-
\frac{\mu_{2}'\lambda_{2}'}{2}+\frac{\mu_{2}'\nu_{2}'}{2}+
\mu_{2}'\phi_{2}')
+Ae^{\nu_{2}-\mu_{2}}=0;
\]
\[
e^{\lambda_{1}-\nu_{1}}(\ddot\phi_{1}+\dot\phi_{1}^{2}-\frac{\dot\nu_{1}\dot\phi_{1}}{2}+
\frac{3}{2}\dot\lambda_{1}\phi_{1})=\gamma;
\]
\[
e^{\nu_{2}-\lambda_{2}}(\phi_{2}''+\phi_{2}'^{2}-\frac{\lambda_{2}'\phi_{2}'}{2}+
\frac{\nu_{2}'\phi_{2}'}{2}+\mu_{2}'\phi_{2}')=\gamma;
\]
\begin{equation}\label{key}
\frac{\nu_{2}'}{2}(\dot\lambda_{1}+\dot\phi_{1})+\phi_{2}'(\frac{\dot\lambda_{1}}{2}-\dot\phi_{1})=0.
\end{equation}

Let us consider following particular cases corresponding to different
ways of disolving of eq. (\ref{key}) separately.

1)$\dot\lambda_{1}=\dot\phi_{1}=0.$ This variant corresponds to well
known and above mentioned Kramer's solution \cite{kram}:
\begin{equation}\label{kramer}
ds^{2}=R^{A-B}dt^{2}-R^{-A-B}dr^{2}-r^{2}R^{1-A-B}d\Omega^{2}
-R^{2B}(dx^{5})^{2},
\end{equation}
where $R=1-\tilde{r}_{g}/r,\ \ A,B,\tilde{r}_{g}$ --- constant of
integration, and $A^{2}+3B^{2}=1$.

2)$\nu_{2}'=0;\ \phi_{2}'=0.$ This case corresponds to 5-D vacuum solutions
of cosmological type, that have been considered in \cite{saz,vlad1,wes5,kok1,
kok2,iv1}
There are a following four solutions:

--- two solutions with a flat space section
\begin{equation}\label{flat1}
ds^{2}=dt^{2}-t(dr^{2}+r^{2}d\Omega^{2})-\frac{1}{t}(dx^{5})^{2};
\end{equation}

\begin{equation}\label{flat2}
ds^{2}=dt^{2}-dr^{2}-r^{2}d\Omega^{2}-t^{2}(dx^{5})^{2}.
\end{equation}

--- solution with 3-D section of constant positive curvature
\begin{equation}\label{close}
ds^{2}=dt^{2}-(b^{2}t^{2}+a)(dr^{2}+\frac{1}{b^{2}}\sinh^{2}(br)d\Omega^{2})
\pm\frac{t^{2}}{b^{2}t^{2}+a}(dx^{5})^{2};
\end{equation}
where $a$ and $b$ are an arbitrary constant of integration;

--- solution with 3-D section of a constant negative curvature,
which can be obtained from proceeding by formal redefenition $b\rightarrow ib$
\begin{equation}\label{open}
ds^{2}=dt^{2}-(a-b^{2}t^{2})(dr^{2}+\frac{1}{b^{2}}\sin^{2}(br)d\Omega^{2})
\pm\frac{t^{2}}{a-b^{2}t^{2}}(dx^{5})^{2};
\end{equation}

3)$\phi_{2}'=-\nu_{2}'/2,\ \ \dot\phi_{1}=-\dot\lambda/4$. Solution has
the following kind
\begin{equation}
ds^{2}=(r^{2}+\frac{a}{r}+1)dt^{2}-4t^{2}dr^{2}
-4t^{2}(r^{2}+\frac{a}{r}+1)r^{2}d\Omega^{2}
-\frac{1}{{\displaystyle t(r^{2}+\frac{a}{r}+1)}}(dx^{5})^{2},
\end{equation}
where $a$ --- constant of integration.

4)$\nu_{2}'=0,\ \dot\phi_{1}=\dot\lambda_{1}/2$. Solution can be put
to the form:
\begin{equation}
ds^{2}=dt^{2}-\frac{t^{2}dr^{2}}{r^{2}+A/r+1}-t^{2}r^{2}d\Omega^{2}+
t^{2}(r^{2}+A/r+1)(dx^{5})^{2},
\end{equation}
where $A$ --- arbitrary constant  of integration.

5)$\dot\lambda_{1}=-\dot\phi_{1},\ \phi_{2}'=0$. It have been founded
one particular solution, which is in turn  particular case of the general A-solution
with $u=-1$.
As for the general case it will be partially analyzed in Appendix.

6)$\dot\phi_{1}=0,\ \ \phi_{2}'=-\nu_{2}'.$ Solution can be put to the form:
\begin{equation}
ds^{2}=\frac{1}{r^{2}}dt^{2}
-\frac{12p^{2}t^{2}}{r^{4}(C_{1}r^{2\sqrt{3}}+
C_{2}r^{-2\sqrt{3}}-2\sqrt{p^{2}+C_{1}C_{2}})^{2}}dr^{2}-
\end{equation}
\[
\frac{p^{2}t^{2}}{r^{2}\sqrt{p^{2}+C_{1}C_{2}}(C_{1}r^{2\sqrt{3}}+C_{2}r^{-2\sqrt{3}}-
2\sqrt{p^{2}+C_{1}C_{2}})}
d\Omega^{2}
-r^{4}(dx^{5})^{2},
\]
where $p,C_{1},C_{2}$ --- arbitrary constants of integration.

7)$\dot\lambda_{1}=0,\ \ \phi_{2}'=\nu_{2}'/2$. Particular solution can be put to the
form:
\begin{equation}
ds^{2}=rdt^{2}-\frac{1}{r^{4}}\frac{1}{(C_{1}r^{\sqrt{3}/2}+C_{2}r^{-\sqrt{3}/2})^{4}}
+
\end{equation}
\[
\frac{d\Omega^{2}}{3C_{1}C_{2}r^{2}(C_{1}r^{\sqrt{3}/2}+C_{2}r^{-\sqrt{3}/2})^{2}}
-t^{2}r(dx^{5})^{2},
\]
where $C_{1},C_{2}$ --- constants of integration.
General case is considering in Appendix.

8)$\alpha=0, \beta=0.$ Solution has the following kind
\begin{equation}
ds^{2}=r^{4}dt^{2}-\frac{St}{r^{4}(C_{1}r^{\sqrt{3}}+C_{2}r^{-\sqrt{3}})^{4}}
dr^{2}
\end{equation}
\[
+\frac{St}{12C_{1}C_{2}r^{2}(C_{1}r^{\sqrt{3}}+C_{2}r^{-\sqrt{3}})^{2}}
-\frac{1}{r^{2}t}(dx^{5})^{2},
\]
where $C_{1},C_{2}$ --- constant of integration.

9)The most general case, to which all above derived solutions are reduced:
$\dot\phi_{1}=(1+\sigma)/(2\sigma-1)\dot\lambda;\ \
\phi_{2}'=4\sigma\nu_{2}'$
It will be considered in Appendix.

There is the following correspondence with results of Wesson,
Liu and Ponce de Leon in \cite{wes2}:

\noindent A-solution (\ref{gen1}) belongs to their class D with $b=const$;\\
A-solution with $u=0$ (\ref{u0}) belongs to class A with $b=const$;\\
A-solution with $u\to\infty$ has been not considered by authors, because of
the wrong signature of this solution;\\
cosmological solutions (\ref{flat1})-(\ref{open}) is related to class B
with $a=1$;\\
case (3) is related to class D under $l=-1/2$;\\
case (4) is related to class C and is its general representator;\\
cases (5) and (6) is related  to class D under $l=-2,0$
correspondingly;\\
case (7) is related to class A with $\lambda=0$. Note, that authors have shown, that in this case
nonstationarity can be excluded by coordinate transformations ;\\
case (8) corresponds to class B under $k=0$;\\
case (9) is related to the most general case D.

\section{ Simulated matter:
5-dimensional approach}

Now we  formulate some general ideas, which can be called
5-dimensional approach to the problem of geometrization of
classical matter.

It is well known that any 5-D vacuum solution after
some mathematical manipulations (1+4-splitting procedure)
can be interpreted as a solution of nonvacuum 4-D Einstein
equations with an effective matter of a definite kind \cite{vlad1,wes4}.
If 5-D metric is independent on fifth coordinate and has no nonzero
components $G_{5\mu}$ then vacuum 5-D equations:
\begin{equation}
{}^{5}\!R_{AB}-\frac{1}{2}G_{AB}{}^{5}\!R=0,
\end{equation}
where $A,B=0,1,2,3,5$, in 4-dimensional representation take the following
form\footnote{In \cite{wes4} another 4-D equations in covariant form
have been obtained. This difference is due to different methods
of 4+1 splitting of starting 5-D equations}\cite{kok1}:
\begin{equation}\label{ein4c}
^{4}R_{\mu\nu}-{1\over2}g_{\mu\nu}{}^{4}R=
(1+2n)\phi_{;\mu;\nu}-(2n^{2}+2n-1)\phi_{,\mu}\phi_{,\nu}
\end{equation}
\[
-g_{\mu\nu}((1+2n)\nabla^{2}\phi + (n^{2}+n+1)(\nabla\phi)^{2});
\]
\begin{equation}\label{scalc}
n\nabla^{2}\phi+ n^{2}(\nabla\phi)^{2}- 1/6{}^{4}R=0,
\end{equation}
    where
$\phi =\ln(\sqrt{-G_{55}}).$
Here parameter $n$ is originated from conformal transformation of starting
4-D metric $\tilde{g}_{\mu\nu}=G_{\mu\nu}-G_{5\mu}G_{5\nu}/G_{55}$,
having the following form:
\[
\tilde{g}_{\mu\nu}=e^{2\phi n}g_{\mu\nu},
\]
where $g_{\mu\nu}$ --- observable metric.
Tensor
\[
T_{\mu\nu}^{(sf)}=(1+2n)\phi_{;\mu;\nu}-(2n^{2}+2n-1)\phi_{,\mu}\phi_{,\nu}
+3n(n+1)g_{\mu\nu}(\nabla\phi)^{2}
\]
where dalambertian is excluded with the help of equation
(\ref{scalc}),
is the energy-momentum tensor of  an effective matter induced by the scalar
field $\phi$. Type of this matter is, in general, arbitrary.
In present article, we assume, that {\it induced matter is anisotropic incoherent
perfect fluid with the some state equation}. Consider separately all
consequences of such hypothesis.

\section{Algebraic type}\label{alg}

Lets analyze algebraic type of symmetric second range tensor $T_{\mu\nu}^{(sf)}.$
From the problem of eigen values and eigen vectors:
\[
T^{(sf)\mu}_{\ \ \ \nu}r^{\nu}=\lambda r^{\mu}
\]
roots of characteristic equation
\[
|T^{(sf)\ \mu}_{\ \ \nu}-\lambda\delta^{\mu}_{\nu}|=0
\]
can be determined. Its type determines algebraic type of tensor
$T^{(sf)}_{\mu\nu}$.
The first consequences of our hypothesis is that {\it algebraic type of
$T^{(sf)}_{\mu\nu}$ must be the same as the anisotropic perfect fluid one.}
Following by Segre notations \cite{Shm}, type of this tensor can be denoted
as the $[1,1,1,1]$,  meaning that this tensor has, in general,  four
different eigen direction with different eigen values.
If normalized eigen vectors are chosen as basis, then $T^{(sf)}_{\mu\nu}$
of such general type can be put to the form:
\[
T^{(sf)}_{\mu\nu}=\epsilon_{0}\lambda_{0}r^{0}_{\mu}r^{0}_{\nu}+
\epsilon_{1}\lambda_{1}r^{1}_{\mu}r^{1}_{\nu}+
\epsilon_{2}\lambda_{2}r^{2}_{\mu}r^{2}_{\nu}+
\epsilon_{3}\lambda_{3}r^{3}_{\mu}r^{3}_{\nu},
\]
where $\lambda_{i},\vec r^{i}$ --- different eigen values and eigen vectors,
$\epsilon_{i}=\pm1$ if corresponding eigen vector timelike or spacelike.
Identification this tensor with physical energy-momentum tensor of
anisotropic perfect fluid gives new restrictions on $\lambda_{i}$ and
$\vec r^{i}$. Naimly: one of the eigen vector, suppose $\vec r^{0}$,
must be timelike, anothers --- spacelike. Then $\lambda_{0}$ is identified
with energy density $\varepsilon:\ \lambda_{0}=\varepsilon$, and
$\lambda_{i}$ are identified with anisotropic pressure: $\lambda_{i}=-p_{i}$
Energy dominancy conditions gives supplement relations:
$\lambda_{0}>0,\ |\lambda_{i}|<\lambda_{0}$

Degeneration (coinsiding) of eigen values leads  to increasing of isotropy
of $T^{(sf)}_{\mu\nu}$. If for example, $\lambda_{i}=\lambda$ for $i=1,2,3$
(algebraic type $[1,(1,1,1)]$)
then tensor $T^{(sf)}_{\mu\nu}$ will correspond to common isotropic
energy-momentum tensor of GR. In section 6 we shall be faced with
tensor of type $[1,1,(1,1)]$ which  is closely connected with the symmetry
of space-time.

If $T^{(sf)}_{\mu\nu}$ has the one nonzero off-diagonal component $T_{01}$
and 4-D metric is diagonal
then the characteristic equation has the following form:
\[
(\lambda^{2}-(T^{0}_{0}+T^{1}_{1})\lambda+(T^{0}_{0}T^{1}_{1}-T^{1}_{0}T^{0}_{1}))(T^{2}_{2}-\lambda)(T^{3}_{3}-\lambda)=0,
\]
roots of which are
\begin{equation}\label{roots}
\lambda_{0,1}=\frac{1}{2}(T^{0}_{0}+T^{1}_{1}\pm\sqrt{(T^{0}_{0}-T^{1}_{1})^{2}
+4T^{0}_{1}T^{1}_{0}});\quad
\lambda_{2}=T^{2}_{2};\ \ \lambda_{3}=T^{3}_{3}.
\end{equation}
Correspondence of roots $\lambda_{0},\lambda_{1}$
to energy density and pressure can be stated from investigations of type of
their eigen vectors: timelike vector is related to  energy density,
spacelike --- to  pressure. For $T^{(sf)}_{\mu\nu}$ of chosen special kind
it is necessary to investigate sign of expression:
\[
1+\frac{g_{11}}{g_{00}}\frac{(T^{1}_{1}-T^{0}_{0}\pm
\sqrt{(T^{1}_{1}-T^{0}_{0})^{2}+4T^{0}_{1}T^{1}_{0}}}{(T^{0}_{1})^{2}}
\]
Eigen vector will be timelike, when this expression positive, and
spacelike when it is negative.

\section{State equation}

After determination of algebraic type, then, if it is suitable,
we should determine what is the type  of obtained perfect fluid
or, in other words, what is the connection between obtained $\epsilon$
and $p$. If tensor $T^{(sf)}_{\mu\nu}$ anisotropic, we shall use the averaged
characteristic $p=(p_{1}+p_{2}+p_{3})/3$. Suppose we have two known
functions $\epsilon$ and $p$ as a function of coordinates.
Parameter
\[
k=\frac{p}{\varepsilon}
\]
in some particular cases can be constant and then will  determine
common linear type of state equation $p=k\varepsilon$. But in general case
$k$ will be function of coordinates and we get "variable state equation".
Let us interpret it by the following way. Assume that perfect fluid with
given $\varepsilon$ and $p$ is the mixture of two noninteracting
comoving
fluids with
linear constant state equation: $p_{1}=k_{1}\varepsilon_{1},\ \
p_{2}=k_{2}\varepsilon_{2}$, where $k_{1},k_{2}$  are constants.
Then their effective energy density will be
$\varepsilon=\varepsilon_{1}+\varepsilon_{2}$
and effective pressure  ---  $p=p_{1}+p_{2}$.
Their effective state equation then will be determined by the
variable parameter $k$:
\begin{equation}\label{konz}
k=\frac{p}{\varepsilon}=\frac{p_{1}+p_{2}}{\varepsilon_{1}+\varepsilon_{2}}=
\frac{k_{1}\varepsilon_{1}+k_{2}\varepsilon_{2}}{\varepsilon_{1}+\varepsilon_{2}}=
\frac{k_{1}+k_{2}n_{21}}{1+n_{21}}.
\end{equation}
Here $n_{21}=\varepsilon_{2}/\varepsilon_{1}$
is relative mass concentration of second fluid to first.
From expression (\ref{konz}) one can get  $n_{21}$ as function of $k$:
\[
n_{21}=\frac{k_{1}-k}{k-k_{2}}.
\]
So, $k(x^{\mu})$ can determine relative distribution of the two
coherent components and its dynamics in the space-time.
Below we shall suppose $k_{1}=0$, $k_{2}=1/3$.
By the same way one can consider $n$ noninteracting coherent
components, but in this case energy densities of $n-2$ components
are arbitrary, and $k$ will  determine  relative concentration
of the remaining two components.

\section{Analysis of obtained solution}

In this section we  apply above discussed ideas to obtained
vacuum solutions. Energy-momentum tensor components have been
calculated with the help of special program in REDUCE.

1){\it Kramer's metric} has been analyzed in details in \cite{wes1,wes3}
under $n=0$. Since this solution is static then energy-momentum tensor
is diagonal. Its components under arbitrary $n$ have the following form:
\[
\begin{array}{l}
T^{0}_{0}=\epsilon=-{\displaystyle \frac{Br_{g}^{2}}{2}}(A(2n+1)+B(2n^{2}+2n-1)){\displaystyle\frac{R^{A-2+B(2n+1)}}{r^{4}}};\\
\\
T^{1}_{1}=-p_{1}=-{\displaystyle\frac{Br_{g}^{2}}{2}}(A(2n+1)+3B(2n^{2}+2n+1)+2(2n+1)(1-2r/r_{g})){\displaystyle\frac{R^{A-2+B(2n+1)}}{r^{4}}};\\
\\
T^{2}_{2}=T^{3}_{3}=-{\displaystyle\frac{Br_{g}^{2}}{2}}(A(2n+1)-B(2n^{2}+2n-1)+(2n+1)(1-2r/r_{g})){\displaystyle\frac{R^{A-2+B(2n+1)}}{r^{4}}};\\
\end{array}
\]
In all cases $T_{\mu\nu}^{(sf)}$
is anisotropic of type $[1,1,(1,1)]$.
Note that particular case $n=-1/2$ corresponds to $k_{1}=1,\ k_{2}=k_{3}=-1$.
It is interesting fact,
that effective state equation, connecting averaged pressure and
energy density, is linear with the coefficient
\[
k=\frac{A(2n+1)-B(10n^{2}+10n+1)}{3(A(2n+1)+B(2n^{2}+2n-1)}.
\]
Case $n=0$ (and also $n=-1$) corresponds to well known result
--- trace of $T^{(sf)}_{\mu\nu}$ is zero \cite{wes3}.

2){\it Vacuum solutions of cosmological type} owing to
homogeneity of 3-D space section give isotropic matter tensor.
Its components for metric (\ref{flat1}) are:
\[
\begin{array}{l}
T^{0}_{0}=\varepsilon={\displaystyle\frac{3(n+1)^{2}}{4t^{n+2}}};\\
\\
T^{1}_{1}=T^{2}_{2}=T^{3}_{3}=-p={\displaystyle\frac{n^{2}-1}{4t^{n+2}}}.
\end{array}
\]
State equation parameter is given by expression:
\[
k=-\frac{n-1}{3(n+1)}.
\]
For the metric (\ref{flat2}) components of effective matter tensor are:
\[
\begin{array}{l}
T^{0}_{0}=\varepsilon=3n^{2}t^{2n-2};\\
\\
T^{1}_{1}=T^{2}_{2}=T^{3}_{3}=-p=n(n+2)t^{2n-2}.
\end{array}
\]
State equation parameter is:
\[
k=-\frac{n+2}{3n}.
\]

For open model (\ref{open}):
\[
\begin{array}{l}
T^{0}_{0}=\varepsilon={\displaystyle\frac{3t^{2n-2}a(an^{2}-
2nb^{2}t^{2}-b^{2}t^{2})}{(a+b^{2}t^{2})^{n+2}}};\\
\\
T^{1}_{1}=T^{2}_{2}=T^{3}_{3}=-p=
{\displaystyle\frac{t^{2n-2}a(an^{2}+2an+2nb^{2}t^{2}+b^{2}t^{2})}
{(a+b^{2}t^{2})^{n+2}}}.
\end{array}
\]
Coefficient $k$ in this case will be variable:
\[
k=-\frac{an(n+2)+(2n+1)t^{2}b^{2}}{3(an^{2}-(2n+1)t^{2}b^{2})}
\]
Expression for relative concentration of dust and radiation is:
\[
n_{21}=\frac{2a(n+1)n}{an(n+2)+(2n+1)t^{2}b^{2}}.
\]
Expressions for closed models can be obtained from opened ones by formal
replacing $b\to ib$.

3){\it Particular case of A-metric under $u=0$}
gives the following effective matter:
\[
\begin{array}{l}
T^{0}_{0}=M(9n^{2}\coth^{2}\sqrt{3}t-n^{2}+2n+2);\\
\\
T^{1}_{1}=M3n((n+2)\coth^{2}\sqrt{3}t-n);\\
\\
T^{1}_{0}=-2\sqrt{3}Mn(n-1)r\coth\sqrt{3}t;\\
\\
T^{0}_{1}=2\sqrt{3}Mn(n-1){\displaystyle\frac{\coth\sqrt{3}t}{r}};\\
\\
T^{2}_{2}=T^{3}_{3}=M((3n^{2}+6n)\coth^{2}\sqrt{3}t-n^{2}-4n-1),
\end{array}
\]
where $M=r^{2n-2}\sinh^{2n}\sqrt{3}t$.
In general case this tensor describes anisotropic two-component perfect
fluid, homogeneously evolved in space. Let us consider the most interesting
cases:

a) $n=0.$ It is easily to see, that tensor $T^{(sf)}_{\mu\nu}$ become
diagonal and  $\varepsilon=2/r^{2},\ \ p_{1}=0,\ \
p_{2}=p_{3}=1/r^{2}$. Corresponding state equation parameters take the
following values: $k_{1}=0,\ k_{2}=k_{3}=1/2$. Average $k=1/3$  and trace of
$T_{\mu\nu}$  under $n=0$ is zero. Trace is zero also under
$n=-1$. Under $n=1$ tensor of effective matter is diagonal too,but
relative concentration of dust and radiation is in this case negative.

b) $n=-1/2.$ For determination of $\varepsilon$ и $p_{1}$ it is necessary to use
formulae (\ref{roots}) from sec.\ref{alg} . Non complicate calculations give
$\varepsilon=p_{1}=p_{2}=p_{3}=(3/4r^{3}\sinh\sqrt{3}t)(3\coth^{2}\sqrt{3}t-1)$
---  stiff matter.

4){\it A-metric} gives the following expressions for
effective matter:
\[
T^{0}_{0}=Mt_{00};\ \ T^{0}_{1}=M\frac{t_{01}t}{r};\ \ T^{1}_{0}=-3M\frac{rt_{01}}{u^{2}t};\ \
\]
\[
T^{1}_{1}=-3M\frac{t_{11}}{u^{2}};\ \ T^{2}_{2}=T^{3}_{3}=-M\frac{3(3-u^{2})t_{22}}{u^{2}},
\]
where
\[
\begin{array}{l}
t_{00}={\displaystyle\frac{6}{u^{2}}}(-nu^{2}+2nu(n-1)+4n^{2}+n+1);\\
\\
t_{01}={\displaystyle\frac{2}{u}}(u^{2}(n^{2}-n)+u(4n^{2}-2n+1)+3(n^{2}-n));\\
\\
t_{11}=2(u^{2}{\displaystyle\frac{(4n^{2}+n+1)}{3}}+2n(n-1)u -3n);\\
\\
t_{22}=t_{33}={\displaystyle\frac{-2n^{2}-2n+1}{3}};\\
\\
M=r^{2n(1+u)-2}t^{2n\frac{{\scriptstyle (3+u)}}{{\scriptstyle u}}-2},
\end{array}
\]
and gives in general linear anisotropic state equation.
Under  $t_{01}=0$ tensor is diagonalized. Zero  $t_{01}$ is obtained,
when parameters  $u$ and $n$  are connected by relation:
\[
u=\frac{-4n^{2}+2n-1\pm\sqrt{4n^{4}+8n^{3}-4n+1}}{2n(n-1)}.
\]
For example, for $n=2,\ u=-1/2$ we get  $k_{1}=-7/18,\ k_{2}=-11/36,\ k=-1$.
Under $n=-1/2\pm\sqrt{3}/2\ \ k_{2}=k_{3}=0,\ k_{1}\neq0$.
Trace is zero, when  $n=0,-1$.

5){\it Metric (3)} generate following effective matter:
\[
\begin{array}{l}
T^{0}_{0}=-{\displaystyle\frac{r^{n-2}}{16t^{n+2}R^{n+2}}}(a^{2}(n^{2}-1)
-8ar^{3}(2n^{2}+3n+1)-12r^{4}(n+1)^{2}\\
-8r^{6}(n^{2}+3n+2));\\
\\
T^{1}_{0}={\displaystyle\frac{r^{n-1}}{8t^{n+3}R^{n+1}}}(n^{2}+3n+2)(a-2r^{3});\\
\\
T^{0}_{1}=-{\displaystyle\frac{r^{n}}{2t^{n+1}R^{n+2}}}(a-2r^{3})(n^{2}+3n+2);\\
\\
T^{1}_{1}=-{\displaystyle\frac{r^{n-2}}{16t^{n+2}R^{n+2}}}(a^{2}(3n^{2}-2n-1)
-4ar^{3}(4n^{2}+8n+3)-4ar(2n+1)\\
-4r^{4}(n^{2}-1)+8r^{6}(n^{2}+3n+2));\\
\\
T^{2}_{2}=T^{3}_{3}=-{\displaystyle\frac{r^{n-2}}{16t^{n+2}R^{n+2}}}(a^{2}(n^{2}-1)
-4ar^{3}(2n^{2}-2n-3)-4r^{4}(n^{2}-2n-2)\\
+4r^{6}(1+2n)),
\end{array}
\]
где $R=a+r^{3}+r$.
Under  $n=-1,-2$ off-diagonal components are vanished. Under $n=-1$
zero value of energy density is get. Under $n=-2$
expression:
\[
\frac{8r^{6}+16r^{4}+20ar^{3}+4ar+8a^{2}}{8r^{6}+20r^{4}+28ar^{3}-4ra-7a^{2}}
\]
in region of its positivity is the relative concentration of dust and radiation

6){\it Metric (4)} gives:
\[
\begin{array}{l}
T^{0}_{0}=-{\displaystyle\frac{t^{2n-2}R^{n-1}n}{4r^{n+3}}}(a^{2}(n+2)-8ar^{3}(1+2n)
-12nr^{4}+8r^{6}(1-n));\\
\\
T^{1}_{0}={\displaystyle\frac{t^{2n-3}R^{n}}{r^{n+2}}}n(n-1)(a-2r^{3});\\
\\
T^{0}_{1}=-{\displaystyle\frac{t^{2n-1}R^{n-1}}{r^{n+1}}}n(n-1)(a-2r^{3});\\
\\
T^{1}_{1}=-{\displaystyle\frac{t^{2n-2}R^{n-1}}{4r^{n+3}}}(a^{2}(3n^{2}+8n+4)+4ar(1+2n)
+4ar^{3}(1-4n^{2})-4r^{4}n(n+2)\\
+8n(n-1)r^{6};\\
T^{2}_{2}=T^{3}_{3}=-{\displaystyle\frac{t^{2n-2}R^{n-1}}{4r^{n+3}}}(a^{2}(n^{2}-2n-2)
-2ar(1+2n)-2ar^{3}(4n^{2}+10n+1)\\-4nr^{4}(n+2)),
\end{array}
\]
where $R=a+r^{3}+r$. Under $n=0,1$ tensor is diagonal.
Under  $n=0$ energy density is vanished. Under $n=1$ we have
anisotropic fluid with $k=-1$.

7){\it Metric (6)} gives:
\vspace{-0.7cm}
\[
\begin{array}{l}
T^{0}_{0}=-{\displaystyle\frac{F^{2}r^{4n+2}(2n^{2}+2n-1)}{6p^{2}t^{2}}};\\
\\
T^{1}_{0}={\displaystyle\frac{F^{2}r^{4n+3}(2n+1)}{6p^{2}t^{3}}};\\
\\
T^{0}_{1}=-{\displaystyle\frac{2r^{4n+1}(2n+1)}{t}};\\
\\
T^{1}_{1}=-{\displaystyle\frac{Fr^{4n+2}((2n+1)F'r+3F(2n^{2}+2n+1)}{6p^{2}t^{2}}};\\
\\
T^{2}_{2}=T^{3}_{3}={\displaystyle\frac{Fr^{4n+2}((1+2n)F'r-2F(2n^{2}+2n-1)}{12p^{2}t^{2}}}.
\end{array}
\]
Here  $F=C_{1}r^{2\sqrt{3}}+C_{2}r^{-2\sqrt{3}}-2\sqrt{p^{2}+C_{1}C_{2}}$.
Trace vanishes under $n=0,1$. Under $n=-1/2$ tensor is diagonalized
and in this case $k_{1}=-1,\ k_{2}=k_{3}=1$.

8){\it Metric (7)} gives:
\[
\begin{array}{l}
T^{0}_{0}=-{\displaystyle\frac{t^{2n-2}r^{n-1}}{4}}(F^{4}r^{3}t^{2}(n^{2}+4n+1)-12n^{2});\\
\\
T^{1}_{0}=-n(n-1)t^{2n-1}r^{n+3}F^{4};\\
\\
T^{0}_{1}=n(n-1)t^{2n-1}r^{n-2};\\
\\
T_{1}^{1}=-{\displaystyle\frac{t^{2n-2}r^{n-1}}{4}}(4(2n+1)F'F^{3}r^{4}t^{2}\\
\\
+3F^{4}r^{3}t^{2}(n+1)^{2}-4n(n+2));\\
\\
T_{2}^{2}=T^{3}_{3}={\displaystyle\frac{t^{2n-2}r^{n-1}}{4}}(2(2n+1)F'F^{3}r^{4}t^{2}\\
\\
-F^{4}r^{3}t^{2}(n^{2}-2n-2)+4n(n+2)),
\end{array}
\]
where $F=C_{1}r^{\sqrt{3}/2}-C_{2}r^{-\sqrt{3}/2}$.
Under $n=0,1$ nondiagonal components vanish. Under $n=0$ we have
anisotropic fluid with $k=1/3$. Under $n=1$ relative concentration of dust
and radiation is
\[
n_{21}=-2\frac{4-F^{4}}{6-F^{4}}
\]
under condition of its positivity: $4\le F^{4}\le 6$.

9){\it Metric (8)}:
\[
\begin{array}{l}
T^{0}_{0}=-{\displaystyle\frac{1}{4t^{n+2}r^{2n+4}}}(4F^{4}r^{6}t(n^{2}-2n-2)-3(n+1)^{2});\\
\\
T^{1}_{0}=-{\displaystyle\frac{F^{4}}{r^{2n-3}t^{n+2}}}(n^{2}+3n+2);\\
\\
T^{0}_{1}={\displaystyle\frac{n^{2}+3n+2}{t^{n+1}r^{2n+5}}};\\
\\
T^{1}_{1}={\displaystyle\frac{8(2n+1)F'F^{3}r^{7}t-12n^{2}F^{4}r^{6}t+n^{2}-1}{4t^{n+2}r^{2n+4}}};\\
\\
T^{2}_{2}=T^{3}_{3}=-{\displaystyle\frac{1}{4t^{n+2}r^{2n+4}}}(4(2n+1)F'F^{3}r^{7}t\\
\\
+4F^{4}r^{6}t(n^{2}+4n+1)-n^{2}+1)\\
\end{array}
\]
Here $F=C_{1}r^{\sqrt{3}}-C_{2}r^{-\sqrt{3}}$.
Tensor diagonalized under $n=-1,-2$. Under $n=-1$ we get anisotropic
fluid with  $k=1/3$, under $n=-2$ expression
\[
-\frac{4(F^{4}r^{6}t-1)}{8F^{4}r^{6}t-3}
\]
describes relative concentration of dust and radiation under
$1/4\le F^{4}r^{6}t\le 3/8$.

\section{Conclusion}

So, the properties of effective matter  can be investigated
in principle for any exact vacuum solution by the proposed way.
The question about physical application of obtained result remained opened.
Probably, it could be applied to some kinds of spherically-symmetric
nonstationar configurations such as stars or elliptic galaxies.
To clear this questions further investigation of properties of the
obtained effective matter is necessary.

Note, that another
approach to the problem of geometrization of matter --- "4-dimensional"
 ---  is possible \cite{kok1}.

\appendix
\section{Analysis of a special cases of
Einstein equations}

In Appendix we  analyze those cases for which exact solutions have not been
founded in apparent kind. By using the special transformations of starting
equations it will be seen that in all  considered cases system of
equations can be reduced to the
Abel's equation of a second kind \cite{zp}. This equation can be integrated in a
quadratures only in some particular cases. So we'll reduce problem  to
the purely
mathematical investigation of equation of a special kind.

Let us start from the case (4) in Sec.\ref{sol}. It characterized that
$\dot\lambda_{1}=-\dot\phi_{1},\ \ \phi_{2}'=0$
System of r-equations (t-equations
can be solved elementary) has the following kind (index "2" is omitted):
\[
\left\{
\begin{array}{l}
\nu''+{\displaystyle\frac{\nu'^{2}}{2}}-{\displaystyle\frac{\nu'\lambda'}{2}}+\nu'\mu'=2\alpha e^{\lambda-\nu};\\
\\
\mu''+{\displaystyle\frac{\mu'^{2}}{2}}+{\displaystyle\frac{\nu''}{2}}+{\displaystyle\frac{\nu'^{2}}{4}}-
{\displaystyle\frac{\lambda'}{2}}\left({\displaystyle\frac{\nu'}{2}}+\mu'\right)=0;\\
\\
\mu''+\mu'^{2}-{\displaystyle\frac{\mu'\lambda'}{2}}+{\displaystyle\frac{\mu'\nu'}{2}}=2Ae^{\lambda-\mu}.
\end{array}
\right.
\]
With choosing special coordinate system: $\mu=2\ln r$, and  denoting
$\nu'=u$ the two last equations can be transformed to the following form:
\[
\frac{u'}{2}+\frac{u^{2}}{4}-\frac{\lambda'u}{4}-\frac{\lambda'}{r}=0;
\]
\[
\frac{2}{r^{2}}-\frac{\lambda'}{r}+\frac{u}{r}=\frac{2Ae^{\lambda}}{r^{2}}.
\]
From this two equations it easily to get its following consequence:
\[
\lambda''r^{2}-3\lambda'r-6Ae^{\lambda}+2A^{2}e^{2\lambda}+4+3A\lambda're^{\lambda}=0;
\]
Introducing new variable $x=\ln r$ equation can be reduced to the form:
\[
\lambda_{xx}-4\lambda_{x}+4-6Ae^{\lambda}+2A^{2}e^{2\lambda}+3A\lambda_{x}e^{\lambda}=0;
\]
Going again to the new variable $z=\lambda$ and new function $\lambda'=p(z)$
last equation can be reduced to the equation of Abel's type:
\[
p'p-p(4-3Ae^{z})+4-6Ae^{z}+2A^{2}e^{2z}=0,
\]
where "\ {}'\ " denote derivative by $z$.
Its the simplest particular solutions are:

1)$p=1-Ae^{z}$ --- is the Shwarzschild solution;

2)$p=0$ --- is the  particular case of A-solution.

General solution with t-dependence has the following form:
\[
ds^{2}=e^{\nu_{2}(r)}dt^{2}-\frac{\alpha}{6}t^{2}e^{\lambda_{2}(r)}dr^{2}-\frac{\alpha}{6}
e^{\mu_{2}(r)}t^{2}d\Omega^{2}
\]
\[
-\frac{1}{t^{4}}(dx^{5})^{2}
\]

For the case (7) ($\dot\lambda_{1}=0,\ \ \phi_{2}'=\nu_{2}'/2$) we have
the following r-system (index "2" is omitted here):

\[
\left\{
\begin{array}{l}
\nu''+\nu'^{2}-{\displaystyle\frac{\nu'\lambda'}{2}}+\nu'\mu'=2\alpha e^{\lambda-\nu};\\
\\
\mu''+{\displaystyle\frac{\mu'^{2}}{2}}-{\displaystyle\frac{\lambda'}{2}}(\mu'+\nu')+\nu''+
{\displaystyle\frac{\nu'^{2}}{2}}=0;\\
\\
\mu''+\mu'^{2}-{\displaystyle\frac{\mu'\lambda'}{2}}+\mu'\nu'-2Ae^{\lambda-\mu}=0.
\end{array}
\right.
\]
Similarly with the previous case suppose $\mu=2\ln r$,
$\nu'=u$.Then two last equations take the following form:
\[
-{\displaystyle\frac{\lambda'}{2}}\left({\displaystyle\frac{2}{r}}+u\right)+u'+\frac{u^{2}}{2}=0;
\]

\[
{\displaystyle\frac{2}{r^{2}}}-{\displaystyle\frac{\lambda'}{r}}+{\displaystyle\frac{2u}{r}}-{\displaystyle\frac{2Ae^{\lambda}}{r^{2}}}=0.
\]
Expressing from the last equation $u$ and its derivative, substituting it
into the first equation and making similar transformations and notations
as in previous case we get the following Abel's type equation
\[
p'p-\frac{1}{4}p^{2}-p(3-2Ae^{x})+3-4Ae^{x}+A^{2}e^{2x}=0.
\]
It has no the solution of the kind $p=k+be^{x}$ as in previous case.
Solution $p=0$ correspond to two considered solutions:
first --- solution of Kasner's type (\ref{flat2}), second --- particular solution of A-solution
under u=0 (\ref{u0}).
General solution with the t-dependence has the following kind:
\[
ds^{2}=e^{\nu_{2}(r)}dt^{2}-e^{\lambda_{2}(r)}dr^{2}-e^{\mu_{2}(r)}d\Omega^{2}
\]
\[
-e^{\nu_{2}(r)}\left\{
\begin{array}{ll}
\sinh^{2}\sqrt{\alpha}t,& \ \alpha>0;\\
\sin^{2}\sqrt{|\alpha|}t,& \ \ \alpha<0;\\
t^{2},&\ \ \ \alpha=0
\end{array}
\right\}(dx^{5})^{2}.
\]

The most general case characterized by the following r-system:
\[
\left\{
\begin{array}{l}
\nu''+{\displaystyle\nu'^{2}}(\sigma+{\displaystyle\frac{1}{2}})-{\displaystyle\frac{\nu'\lambda'}{2}}+\nu'\mu'=
2\alpha e^{\lambda-\nu};\\
\\
\mu''+{\displaystyle\frac{\mu'^{2}}{2}}+\nu''({\displaystyle\frac{1}{2}}+\sigma)+\nu'^{2}({\displaystyle\frac{1}{4}}+
\sigma^{2})
-{\displaystyle\frac{\lambda'}{2}}(\mu'+\nu'({\displaystyle\frac{1}{2}}+\sigma))=\beta e^{\lambda-\nu};\\
\\
\beta-{\displaystyle\frac{e^{\nu-\lambda}}{2}}(\mu''+\mu'^{2}-{\displaystyle\frac{\mu'\lambda'}{2}}+
\mu'\nu'({\displaystyle\frac{1}{2}}+\sigma))
+Ae^{\nu-\mu}=0,
\end{array}
\right.
\]
and following conditions on a separating constant: $(1+\sigma)/(2\sigma-1)\beta=\sigma\alpha.$
Lets take coordinate system where $\nu=2\ln r$.
Then first and second equations take the following kinds:
\[
\frac{4\sigma}{r^{2}}-\frac{\lambda'}{r}+\frac{2\mu'}{r}=\frac{2\alpha e^{\lambda}}{r^{2}};
\]

\[
\mu''+\frac{\mu'^{2}}{2}+\frac{1}{r^{2}}(4\sigma^{2}-2\sigma)-
\frac{\lambda'}{2}(\mu'+\frac{2}{r}(\frac{1}{2}+\sigma))=\frac{\beta e^{\lambda}}{r^{2}}.
\]

Expressing from the first equation $\mu'$, inserting it into second and
transforming last by the similar manner as in previous cases we get
the following equation of Abel's type:
\[
p'p-p(2+2\sigma-\alpha e^{x})-\frac{p^{2}}{4}+12\sigma^{2}
\]
\[
-(\alpha(1+2\sigma)+\beta)e^{x}+\frac{\alpha^{2}}{2}e^{2x}=0.
\]
There is no solution of kind $p=k+be^{x}$ as in first considered here case.
Particular solution $p=0$ is the considered solution (\ref{gen1}).
General form of solution with the t-dependence is
\[
ds^{2}=e^{\nu_{2}(r)}dt^{2}-t^{2}e^{\lambda_{2}(r)}dr^{2}-t^{2}e^{\mu_{2}(r)}d\Omega^{2}-
t^{\frac{{\scriptstyle 4(1+\sigma)}}{{\scriptstyle 2\sigma-1}}}e^{8\sigma\nu_{2}(r)}(dx^{5})^{2}.
\]

%

%
%

\end{document}